# Diverse high-Chern-number quantum anomalous Hall insulators in twisted rhombohedral graphene

**Authors:** Naitian Liu[1], Zhangyuan Chen[1], Jing Ding[1], Wenqiang Zhou[1], Hanxiao Xiang[1], Xinjie Fang[1], Linfeng Wu[1], Xiaowan Zhan[2], Le Zhang[1,3], Qianmei Chen[2], Kenji Watanabe[4], Takashi Taniguchi[5], Na Xin[2*], Shuigang Xu[1,3*]

**Affiliations:**
[1] Key Laboratory for Quantum Materials of Zhejiang Province, Department of Physics, School of Science, Westlake University, Hangzhou 310030, China
[2] Department of Chemistry, Zhejiang University, Hangzhou 310058, China
[3] Institute of Natural Sciences, Westlake Institute for Advanced Study, Hangzhou 310024, China
[4] Research Center for Electronic and Optical Materials, National Institute for Materials Science, Tsukuba 305-0044, Japan
[5] Research Center for Materials Nanoarchitectonics, National Institute for Materials Science, Tsukuba 305-0044, Japan

[*]Corresponding authors. Email: na.xin@zju.edu.cn (N.X.); xushuigang@westlake.edu.cn (S.X.)

**Abstract:**
Quantum anomalous Hall (QAH) insulators with high Chern number ($C$) enables multiple dissipationless edge channels for low-power-consumption electronics. We report the realization of multiple high-$C$ QAH insulators including $C = 3, 5, 6$, and $7$ in twisted monolayer-rhombohedral pentalayer graphene. In twist angles of approximately $1.40°$, we observe QAH effect with $C = 5$ at a filling of one electron per moiré unit cell, persisting up to 2 Kelvin. Furthermore, incommensurate QAH insulators with $C = 5, 6$, and $7$ emerge at partial fillings. In twist angles of $0.89°$, Chern insulators with $C = 3$ and $C = 6$ appear at fillings of two and three electrons, respectively. Our findings establish twisted rhombohedral multilayer graphene as a highly tunable platform for multichannel, dissipationless electronics and for the exploration of exotic quantum Hall states beyond traditional Landau level paradigm.

**Main text:**

The quantum anomalous Hall (QAH) effect is a lattice analogue of the quantum Hall effect, exhibiting quantized Hall resistance of $h/(Ce^2)$ and vanishing longitudinal resistance in the absence of an external magnetic field $B$. Here, $h$ is Planck's constant, $e$ is the elementary charge, and $C$ is the Chern number—an integer that characterizes the number of chiral edge channels (*1*). Realizing high-Chern-number ($|C| > 1$) QAH states not only enables multiple dissipationless edge channels for low-power-consumption electronic applications but also enriches the landscape of correlated topological phases in fractional Chern insulators (FCI) regime (*2-5*). In topological bands with $|C| = 1$, FCI can be continuously connected to conventional fractional quantum Hall states in continuum Landau levels, which always have $|C| = 1$. By contrast, fractional quantum Hall states in bands with $|C| > 1$ have no direct Landau level counterpart, opening the door to fundamentally new quantum states of matter. While QAH insulators with $|C| = 1$ have been realized in several systems (*6-12*), experimental reports of high-Chern-number QAH insulators remain rare (*13-20*). The reported highest-$C$ QAH insulators, such as spin-orbit proximitized multilayer graphene (*13*) and multilayer magnetically doped topological insulators (*14*), have not shown signatures of FCI, likely due to the lack of well isolated flat bands needed to stabilize fractional fillings.

Rhombohedral multilayer graphene is a compelling platform for engineering strongly correlated and topological physics due to its layer-number-dependent flat surface bands and large Berry curvature near charge neutrality point (*21-23*). When crystallographically aligned with hexagonal boron nitride (h-BN), moiré superlattices form and isolate low-energy flat bands with nonzero valley Chern numbers (*16, 24-26*). Consequently, abundant topological states have been observed in this system, including both integer and fractional QAH effects (*12, 18, 20, 27-29*). The electronic properties of these moiré superlattices are sensitive to the twist angle $\theta$, which controls the moiré wavelength, the band width and correlation strength (*24, 25, 30*). However, in graphene/h-BN systems, the achievable moiré wavelength is limited (~15 nm at $\theta = 0°$), and twist angle control remains a challenge.

An alternative strategy of constructing moiré superlattice is stacking two atomically thin crystals of the same material with a relatively small twist angle, such as twisted Bernal-stacked graphene (*31-34*). Compared with graphene/h-BN superlattices, twisted graphene/graphene structures offer better twist angle control and larger tunability of moiré periodicity via well-developed "cut-and-stack" techniques (*35-38*). In such systems, the valley Chern number of low-energy bands depends on the number of graphene layers, suggesting that high-Chern-number QAH states could emerge in twisted rhombohedral multilayer graphene with sufficiently large layer numbers (*39, 40*).

Here, we report the observation of diverse high-Chern-number QAH insulators in twisted monolayer-rhombohedral pentalayer graphene moiré superlattices. We find the electric field and twist angle provide effective means of tuning the Chern number of the resulting topological states. In devices with a twist angle of $\theta \approx 1.40°$, we observe QAH insulators with a quantized Hall resistance of $R_{xy} = h/5e^2$, accompanied by pronounced magnetic hysteresis at a filling of one electron per moiré unit cell ($\nu = 1$). Intriguingly, incommensurate QAH states with $C = 5, 6$, and $7$, which are tunable by electric fields, also emerge at higher fillings ($\nu > 1$). In devices with $\theta \approx 0.89°$, high-Chern-number Chern insulators appear at fillings of two electrons ($\nu = 2$) and three electrons ($\nu = 3$) per moiré unit cell with $C = 3$ and $C = 6$, respectively.

**Phase diagram of device D1 with a twist angle of $1.40°$**

Our device structure is illustrated in Fig. 1A (*41, 42*), where a moiré superlattice forms at the interface between two graphene sheets with a small twist angle $\theta$. We use monolayer and rhombohedral pentalayer graphene as the building blocks. On the one hand, monolayer graphene ensures that moiré hybridization primarily occurs at the surface, making surface transport dominant in the low-energy band. On the other hand, rhombohedral pentalayer graphene has been demonstrated as a wonderful playground for exploring correlated topological states, as it balances the tradeoff between the flatness of the surface band and trigonal warping effect (*12, 43, 44*). The dual-gate structure allows independent control of carrier density $n$ and displacement field $D$. We have measured four devices with varying twist angles. In the main text, we present data from devices D1, D2 and D3, with data from devices D4 and additional characterizations included in Supplementary Materials (*41*).

Figure 1B displays the full $n - D$ map of the longitudinal resistance $R_{xx}$, measured in a dilution refrigerator at $B = 0$ T and the temperature of $T = 50$ mK for device D1. The twist angle is determined to be $\theta \approx 1.40°$ from electrical transport measurements, corresponding to a moiré period of ~10.1 nm. The phase diagram exhibits pronounced asymmetry with respect to $D$, attributed to the structure asymmetry of the sample. Besides the single-particle gaps at charge neutrality and full band filling ($\nu = 4$), multiple symmetry-broken states appear at $\nu = 1$ and $2$ on the $D < 0$ side, arising from strong correlations in the flat bands. Intriguingly, these correlated states emerge when electrons are polarized toward the moiréless interface under large negative $D$ (see fig. S3) (*41*), where the single-particle gap expected at $\nu = 4$ is absent at the same $D$. This unconventional phase diagram resembles that of rhombohedral multilayer graphene/h-BN moiré systems, triggering the discussions of the role of strong correlations in opening topological gaps (*12, 20, 27-29, 45-50*). Nevertheless, unlike graphene/h-BN moiré systems, we do not observe correlated states (e.g., at $\nu = 1, 2, 3$) on the strong moiré side ($D > 0$) at zero magnetic field; these states only emerge under finite magnetic fields (see fig. S6) (*41*).

As shown in Fig. 1B, notably, the $\nu = 1$ state under $D < 0$ exhibits a pronounced dip in $R_{xx}$, which is the main focus of this study. To unveil the nature of this state, we further simultaneously measured $R_{xx}$ and $R_{xy}$ near $\nu = 1$ in detail. Figure 1C and 1D show the fine maps of $R_{xx}$ and $R_{xy}$, zooming in on the boxed region from Fig. 1B. To suppress residual mixing between $R_{xx}$ and $R_{xy}$ and reduce zero-field magnetic fluctuations, $R_{xx}$ ($R_{xy}$) was symmetrized (anti-symmetrized) with respect to small fields ($B = \pm 0.1$ T). Strikingly, an unusually large $R_{xy}$, concomitant with a nearly vanishing $R_{xx}$, emerges near $\nu = 1$, along with a nearby wing-like feature extending toward $\nu \approx 1.5$. The large Hall angle in this region points to the emergence of a nontrivial topological phase.

**QAH insulators with $C = 5$ at $\nu = 1$**

To further probe the topology nature of the $\nu = 1$ state, we performed magnetic hysteresis measurements at a fixed $D = -0.620$ V nm$^{-1}$ by sweeping $B$ forward and backward as shown in Fig. 2A. Remarkably, a sharp hysteretic loop is observed, indicating spontaneous time-reversal symmetry breaking. The Hall resistance $R_{xy}$ switches sign abruptly at a coercive field $B_c \approx 13$ mT, forming a parallelogram-like loop characteristic of a ferromagnetic state. This anomalous Hall signal gradually weakens with increasing $T$, with a Curie temperature around 7 K (see Fig. 2B and fig. S9) (*41*). Notably, $R_{xy}$ is quantized at $h/5e^2 \approx 5.16$ kΩ, persisting to zero external $B$, accompanied by a vanishing $R_{xx}$ below 300 Ω. The quantized plateau persists up to approximately 2.0 K (see Fig. 2B and fig. S9) (*41*). We estimate the energy gap of the $\nu = 1$ state by fitting the temperature dependence of $R_{xx}$ ($R_{xy}$) to an Arrhenius form $R_{xx} \sim e^{-\Delta/2k_BT}$ ($\delta R_{xy} = \frac{h}{5e^2} - |R_{xy}| \propto e^{-\Delta/k_BT}$), where $\Delta$ is the energy gap and $k_B$ is Boltzmann constant. It yields a gap size of $\Delta \approx 16.31$ K (14.33 K). The quantized Hall plateau is also observed by sweeping $n$ at a fixed $D = -0.620$V nm$^{-1}$ (Fig. 2C).

All the above features confirm that the $\nu = 1$ state is a QAH insulator with a Chern number of $C = 5$. This assignment is further supported by Landau fan diagrams derived from $R_{xx}$ and $R_{xy}$ as functions of carrier density $n$ and out-of-plane magnetic field $B$. In QAH insulators, the slope of the topologically gapped state's trajectory in the $(n, B)$ map follows the Streda formula $\frac{\partial n}{\partial B} = C\frac{e}{h}$, from which one can extract the Chern number $C$. As shown in Fig. 2D and 2E, we observe remarkably dispersive fan diagrams emanating from $\nu = 1$, with $R_{xy}$ switching its sign but retaining the same magnitude across zero field. The slopes of the dispersion marked by the dashed lines in either Fig. 2D or Fig. 2E yield a Chern number of $C = 5$ according to the Streda formula, excellently agreeing with the quantized value of $R_{xy}$. We also observe competing Chern states with opposite signs at finite $B$ (see fig. S8) (*41*), reminiscent of similar observations in other correlated moiré systems (*20, 51*).

Notably, the observed $C = 5$ QAH state in twisted monolayer-rhombohedral pentalayer graphene moiré superlattices in this study is apparently distinct from the $C = 1$ QAH states previously reported in rhombohedral pentalayer graphene/h-BN moiré superlattices (*12*), indicating that not only moiré potential but also interlayer coupling critically influence the topological band structure. By utilizing twist graphene homostructures, we recover the high-Chern-number Chern band arising from the large Berry curvature in pentalayer graphene. Our results are consistent with theoretical predictions for twisted rhombohedral graphene homostructures, in which the Chern number follows $C = n + m - 1$ for a twisted $n$-layer on $m$-layer rhombohedral graphene near the magic angle (*52*).

**Incommensurate QAH insulators with $C = 5, 6,$ and $7$**

Intriguingly, we find that the QAH states at $\nu = 1$ extends continuously into the range $1 < \nu < 1.5$. As shown in Fig. 1C, local minima in $R_{xx}$ trace an additional branch over this broad filling range, accompanied by anomalous $R_{xy}$. This feature is reproducible in our device D2 with a twist angle of $\theta \approx 1.32°$, as shown in Fig. 3A. In contrast to valley-polarized metallic states (*43*), these states exhibit topologically nontrivial characteristics, as revealed by their unconventional behaviors in the Landau fan diagrams. Figure 3C display $R_{xy}$ as functions of $\nu$ and $B$ at a fixed $D = -0.612$ V nm$^{-1}$. In addition to the established QAH insulators emanating from $\nu = 1$, the dispersive Landau fan extends a large region of $\nu$ beyond $\nu = 1$ (up to $\nu = 1.3$), resembling the extended QAH states in rhombohedral graphene/h-BN moiré superlattices (*27*). The anomalous Hall effect in the region of $1 < \nu < 1.5$ is further corroborated by remarkable magnetic hysteresis in $R_{xy}$ as shown in Fig. 3F-H, indicating spontaneous time-reversal symmetry breaking. At $\nu = 1.13$, $R_{xy}$ is quantized at $h/5e^2$ down to $B \approx 25$ mT (Fig. 3F), which is consistent with a Chern number of $C = 5$ using the Streda formula in Fig. 3C, matching that of the adjacent $\nu = 1$ state.

Moreover, we find that the Chern number of incommensurate QAH insulators can deviate from that of the parent $\nu = 1$ state with further increasing $|D|$. Figure 3B demonstrates the coexistence of multiple QAH states with $C = 5, 6,$ and $7$, achievable by tuning $D$. Of particular interest, these QAH states appear at filling factors incommensurate with the moiré superlattices. Figure 3C-3E illustrate the evolution of these incommensurate QAH states as $D$. Strikingly, their zero-field intercept filling factors shift with increasing $|D|$. Figure 3D and 3E show the Landau fan diagrams at fixed $D = -0.632$ V nm$^{-1}$ and -0.651 V nm$^{-1}$, respectively, demonstrating new QAH states of $C = 6$ ($C = 7$) emanating from $\nu = 1.28$ ($\nu = 1.40$) as determined via the Streda formula. The corresponding magnetic hysteresis near $B = 0$ further confirms their QAH nature (Fig. 3G,H). Specifically, at $\nu = 1.28$ and $D = -0.632$ V nm$^{-1}$, $R_{xy}$ is quantized at $h/6e^2 \approx 4.3$ k$\Omega$, persisting down to $B = 5$ mT and reaches 96% of the quantized value at $B = 0$. Similarly, at $\nu = 1.40$ and $D = -0.651$ V nm$^{-1}$, $R_{xy}$ approaches $h/7e^2 \approx 3.7$ k$\Omega$, persisting down to $B \approx 50$ mT, and reaches 91% of the quantized value at $B = 0$.

The emergence of incommensurate QAH insulators in our system is reminiscent of recent observations of anomalous Hall crystals in rhombohedral heptalayer graphene/h-BN superlattices (*50*), where similar features occur between $\nu = 1$ and $\nu = 2$. The incommensurate QAH states in this study exhibit both the same Chern number ($C = 5$) and distinct Chern number ($C = 6, 7$) relative to their parent bands, depending on $D$. Two possible underlying mechanisms could account for these states. First, they may originate from the formation of Wigner crystals with excess electrons superimposed on a $\nu = 1$ Chern state, akin to re-entrant quantum Hall states (*27, 53-55*). In this scenario, excess electrons condense into Wigner crystals due to the large Wigner-Seitz radius arising from the flat band's large effective mass, providing a topologically trivial background (*56, 57*). Meanwhile, the topologically nontrivial features between $1 < \nu < 1.5$ inherit from the $\nu = 1$ state with $C = 5$. Alternatively, a more exotic interpretation involves the direct formation of anomalous Hall crystals via strong-correlation-driven spontaneous time-reversal and continuous translational symmetry breakings simultaneously (*45-49*), resembling the scenario in rhombohedral heptalayer graphene/h-BN systems (*50*). Given that the Chern number of the incommensurate QAH states here is tunable by $D$ and can be distinct from their parent states, the latter scenario—anomalous Hall crystal formation driven by strong correlations in the flat Chern band of twisted rhombohedral graphene—appears most plausible.

**High-Chern-number Chern insulators in a small-twist-angle device**
An additional advantage of twist graphene homostructures lies in the possibility of controlling over the twist angle, enabled by well-developed "cut-and-stack" technique (*35-38*). Previous theoretical calculations suggest the valley Chern number in twisted graphene systems strongly depends on the twist angle, which provides additional control knobs for Chern-state engineering besides the electric fields (*39, 40, 58, 59*). Here, we experimentally reveal that the Chern states in twisted monolayer-rhombohedral pentalayer graphene are highly tunable by twist angle, enabling access to diverse high-Chern-number Chern states. Figure 4 presents the phase diagram of device D3, which has a twist angle of $\theta \approx 0.89°$ (fig. S11) (*41*). The corresponding moiré wavelength reaches ~15.8 nm, also confirmed by Brown-Zak oscillations (fig. S4) (*41*), exceeding the maximum moiré wavelength in graphene/h-BN moiré systems. Unlike devices D1 and D2, device D3 exhibits multiple resistance peaks at full fillings of moiré bands ($\nu = 4, 8, 12$) under $D > 0$ (fig. S11) (*41*). Nevertheless, we find the topological states still emerge under $D < 0$ when electrons are polarized toward the moiréless interface.

Figure 4A and 4B display the symmetrized $R_{xx}$ and antisymmetrized $R_{xy}$ maps, respectively, acquired at $T = 50$ mK and $B = \pm 0.1$ T. In contrast to devices D1 and D2, device D3 exhibits topological nontrivial features at $\nu = 2$ and $\nu = 3$. The Landau fan diagrams in Fig. 4C and 4D (taken at $D = -0.743$ V nm$^{-1}$) manifest dispersive behaviors whose slopes (indicated by the dashed line) yield Chern numbers of $C = 3$ at $\nu = 2$ and $C = 6$ at $\nu = 3$, extracted via the Streda formula. These assignments are further supported by the observations of magnetic hysteresis loops. At $\nu = 2$, $R_{xy}$ approaches the quantized value $h/3e^2$ at $B \approx 0.12$ T (Fig. 4E), while at $\nu = 3$, $R_{xy}$ reaches 93% of the quantized value $h/6e^2$ down to $B \approx 10$ mT (Fig. 4F).

Notably, a nontrivial Chern state may also exist at $\nu = 1$, as suggested by the enhanced $R_{xy}$ (Fig. 4B) and weak magnetic hysteresis near $\nu = 1$ (fig. S11) (*41*). However, the presence of a highly resistive state spanning $0 < \nu < 1$ obscures the precise extraction of the Chern number at $\nu = 1$. Such ambiguity is likely due to moiré disorder, which becomes more pronounced at smaller twist angles and low carrier densities (*60*). Further improvements in sample quality and twist angle homogeneity will be necessary to resolve the nature of the $\nu = 1$ state definitively.

**Discussion and outlook**
Our findings of high-Chern-number QAH insulators in twisted rhombohedral graphene represent the highest Chern numbers ($C = 5, 6,$ and $7$) reported to date. The presence of multiple chiral edge channels in these high-Chern-number phases offers potential advantages for low-dissipation transport, including reduced contact resistance and improved energy efficiency, thereby enhancing the performance of quantum devices. Unlike previously reported high-Chern-number QAH insulators in systems without moiré superlattices (*13, 14*), our system features an isolated flat Chern band. The large Chern number arises from the giant momentum-space Berry curvature in the low-energy surface band of rhombohedral pentalayer graphene (*21, 61*). The moiré superlattices isolate this surface band either through zone folding or strong correlations, giving rise to topological bands with valley-contrasting Chern numbers (*40, 45*). The sufficiently flat band enhances Coulomb interactions which in turn spontaneously lift the spin/valley flavors degeneracy (*62-64*). At $\nu = 1$, a single spin- and valley-polarized Chern band becomes populated, analogous to behavior seen in other moiré systems such as twisted bilayer graphene aligned with h-BN and twisted MoTe$_2$ (*8, 10, 11*), but with distinct Berry curvature due to the rhombohedral pentalayer graphene used. The spontaneous valley symmetry breaking from Coulomb interactions gives rise to large valley-contrasting orbital magnetizations associated with quantized anomalous Hall effect (*39*). The QAH at other fillings (such as $1 < \nu < 1.5, \nu = 2,$ and $\nu = 3$) may originate from more exotic mechanisms, requiring further experimental and theoretical studies.

The realization of multiple high-Chern-number QAH states within an isolated flat band opens a promising avenue for the exploration of unconventional FCI without Landau level analogues (*2-5*). Since the Chern number in our system is intrinsically linked to the layer number of rhombohedral multilayer graphene (*52*), it is possible to achieve an even higher Chern number in a thicker rhombohedral graphene, further enriching the landscape of possible fractionized phases. In contrast to rhombohedral graphene/h-BN superlattices, twisted graphene homostructures offer deterministic tunability of the twist angle—and thus the moiré wavelength—through fabrication. This tunability, combined with electric field and carrier density control, allows systematic access to a wide range of correlated and topological phases. Such versatility establishes twisted rhombohedral graphene as an ideal platform for engineering novel QAH and fractional QAH states, with potential realizations of bosonic QAH and non-Abelian anyons (*3, 5,*

*65*). Our findings thus provide a robust experimental foundation and roadmap for future exploration of strongly correlated topological phases in engineered flat bands.

**References and Notes**
1. F. D. M. Haldane, Model for a quantum Hall effect without Landau levels: condensed-matter realization of the "parity anomaly". *Phys. Rev. Lett.* **61**, 2015-2018 (1988).
2. Y.-F. Wang, H. Yao, C.-D. Gong, D. N. Sheng, Fractional quantum Hall effect in topological flat bands with Chern number two. *Phys. Rev. B* **86**, 201101 (2012).
3. S. Yang, Z.-C. Gu, K. Sun, S. Das Sarma, Topological flat band models with arbitrary Chern numbers. *Phys. Rev. B* **86**, 241112 (2012).
4. Z. Liu, E. J. Bergholtz, H. Fan, A. M. Läuchli, Fractional Chern insulators in topological flat bands with higher Chern number. *Phys. Rev. Lett.* **109**, 186805 (2012).
5. G. Möller, N. R. Cooper, Fractional Chern insulators in Harper-Hofstadter bands with higher Chern number. *Phys. Rev. Lett.* **115**, 126401 (2015).
6. C.-Z. Chang *et al.*, Experimental observation of the quantum anomalous Hall effect in a magnetic topological insulator. *Science* **340**, 167-170 (2013).
7. Y. Deng *et al.*, Quantum anomalous Hall effect in intrinsic magnetic topological insulator $MnBi_2Te_4$. *Science* **367**, 895-900 (2020).
8. M. Serlin *et al.*, Intrinsic quantized anomalous Hall effect in a moiré heterostructure. *Science* **367**, 900-903 (2020).
9. T. Li *et al.*, Quantum anomalous Hall effect from intertwined moiré bands. *Nature* **600**, 641-646 (2021).
10. F. Xu *et al.*, Observation of integer and fractional quantum anomalous Hall effects in twisted bilayer $MoTe_2$. *Phys. Rev. X* **13**, 031037 (2023).
11. H. Park *et al.*, Observation of fractionally quantized anomalous Hall effect. *Nature* **622**, 74-79 (2023).
12. Z. Lu *et al.*, Fractional quantum anomalous Hall effect in multilayer graphene. *Nature* **626**, 759-764 (2024).
13. T. Han *et al.*, Large quantum anomalous Hall effect in spin-orbit proximitized rhombohedral graphene. *Science* **384**, 647-651 (2024).
14. Y.-F. Zhao *et al.*, Tuning the Chern number in quantum anomalous Hall insulators. *Nature* **588**, 419-423 (2020).
15. J. Ge *et al.*, High-Chern-number and high-temperature quantum Hall effect without Landau levels. *Natl. Sci. Rev.* **7**, 1280-1287 (2020).
16. G. Chen *et al.*, Tunable correlated Chern insulator and ferromagnetism in a moiré superlattice. *Nature* **579**, 56-61 (2020).
17. H. Polshyn *et al.*, Electrical switching of magnetic order in an orbital Chern insulator. *Nature* **588**, 66-70 (2020).
18. Y. Choi *et al.*, Superconductivity and quantized anomalous Hall effect in rhombohedral graphene. *Nature* **639**, 342–347 (2025).
19. Y. Sha *et al.*, Observation of a Chern insulator in crystalline ABCA-tetralayer graphene with spin-orbit coupling. *Science* **384**, 414-419 (2024).

20. J. Ding *et al.*, Electric-field switchable chirality in rhombohedral graphene Chern insulators stabilized by tungsten diselenide. *Phys. Rev. X* **15**, 011052 (2025).
21. F. Zhang, J. Jung, G. A. Fiete, Q. Niu, A. H. MacDonald, Spontaneous quantum Hall states in chirally stacked few-layer graphene systems. *Phys. Rev. Lett.* **106**, 156801 (2011).
22. M. Koshino, Interlayer screening effect in graphene multilayers with ABA and ABC stacking. *Phys. Rev. B* **81**, 125304 (2010).
23. N. B. Kopnin, M. Ijas, A. Harju, T. T. Heikkila, High-temperature surface superconductivity in rhombohedral graphite. *Phys. Rev. B* **87**, 140503(R) (2013).
24. B. L. Chittari, G. Chen, Y. Zhang, F. Wang, J. Jung, Gate-tunable topological flat bands in trilayer graphene boron-nitride moiré superlattices. *Phys. Rev. Lett.* **122**, 016401 (2019).
25. Y. Park, Y. Kim, B. L. Chittari, J. Jung, Topological flat bands in rhombohedral tetralayer and multilayer graphene on hexagonal boron nitride moiré superlattices. *Phys. Rev. B* **108**, 155406 (2023).
26. W. Zhou *et al.*, Layer-polarized ferromagnetism in rhombohedral multilayer graphene. *Nat. Commun.* **15**, 2597 (2024).
27. Z. Lu *et al.*, Extended quantum anomalous Hall states in graphene/hBN moiré superlattices. *Nature* **637**, 1090–1095 (2025).
28. J. Xie *et al.*, Tunable fractional Chern insulators in rhombohedral graphene superlattices. *Nat. Mater.* **24**, 1042-1048 (2025).
29. D. Waters *et al.*, Chern insulators at integer and fractional filling in moiré pentalayer graphene. *Phys. Rev. X* **15**, 011045 (2025).
30. Z. Dong, A. S. Patri, T. Senthil, Stability of anomalous Hall crystals in multilayer rhombohedral graphene. *Phys. Rev. B* **110**, 205130 (2024).
31. Y. Cao *et al.*, Unconventional superconductivity in magic-angle graphene superlattices. *Nature* **556**, 43-50 (2018).
32. S. G. Xu *et al.*, Tunable van Hove singularities and correlated states in twisted monolayer-bilayer graphene. *Nat. Phys.* **17**, 619-626 (2021).
33. C. Shen *et al.*, Correlated states in twisted double bilayer graphene. *Nat. Phys.* **16**, 520-525 (2020).
34. X. Liu *et al.*, Tunable spin-polarized correlated states in twisted double bilayer graphene. *Nature* **583**, 221-225 (2020).
35. Y. Cao *et al.*, Superlattice-induced insulating states and valley-protected orbits in twisted bilayer graphene. *Phys. Rev. Lett.* **117**, 116804 (2016).
36. K. Kim *et al.*, van der Waals heterostructures with high accuracy rotational alignment. *Nano Lett.* **16**, 1989-1995 (2016).
37. Y. Saito, J. Ge, K. Watanabe, T. Taniguchi, A. F. Young, Independent superconductors and correlated insulators in twisted bilayer graphene. *Nat. Phys.* **16**, 926-930 (2020).
38. J. Díez-Mérida *et al.*, High-yield fabrication of bubble-free magic-angle twisted bilayer graphene devices with high twist-angle homogeneity. *Newton* **1**, 100007 (2025).
39. J. Liu, Z. Ma, J. Gao, X. Dai, Quantum valley Hall effect, orbital magnetism, and

anomalous Hall effect in twisted multilayer graphene systems. *Phys. Rev. X* **9**, 031021 (2019).

40. Y.-H. Zhang, D. Mao, Y. Cao, P. Jarillo-Herrero, T. Senthil, Nearly flat Chern bands in moiré superlattices. *Phys. Rev. B* **99**, 075127 (2019).
41. Materials and methods are available as supplementary materials.
42. Z. Feng *et al.*, Rapid infrared imaging of rhombohedral graphene. *Phys. Rev. Applied* **23**, 034012 (2025).
43. T. Han *et al.*, Orbital multiferroicity in pentalayer rhombohedral graphene. *Nature* **623**, 41-47 (2023).
44. T. Han *et al.*, Correlated insulator and Chern insulators in pentalayer rhombohedral-stacked graphene. *Nat. Nanotechnol.* **19**, 181-187 (2024).
45. T. Tan, T. Devakul, Parent Berry curvature and the ideal anomalous Hall crystal. *Phys. Rev. X* **14**, 041040 (2024).
46. Z. Dong, A. S. Patri, T. Senthil, Theory of quantum anomalous Hall phases in pentalayer rhombohedral graphene moiré structures. *Phys. Rev. Lett.* **133**, 206502 (2024).
47. J. Dong *et al.*, Anomalous Hall crystals in rhombohedral multilayer graphene. I. Interaction-driven Chern bands and fractional quantum Hall states at zero magnetic field. *Phys. Rev. Lett.* **133**, 206503 (2024).
48. B. Zhou, H. Yang, Y.-H. Zhang, Fractional quantum anomalous Hall effect in rhombohedral multilayer graphene in the moiréless limit. *Phys. Rev. Lett.* **133**, 206504 (2024).
49. D. N. Sheng, A. P. Reddy, A. Abouelkomsan, E. J. Bergholtz, L. Fu, Quantum anomalous Hall crystal at fractional filling of moiré superlattices. *Phys. Rev. Lett.* **133**, 066601 (2024).
50. H. Xiang *et al.*, Continuously tunable anomalous Hall crystals in rhombohedral heptalayer graphene. arXiv:2502.18031 (2025).
51. R. Su *et al.*, Moiré-driven topological electronic crystals in twisted graphene. *Nature* **637**, 1084–1089 (2025).
52. P. J. Ledwith, A. Vishwanath, E. Khalaf, Family of ideal Chern flatbands with arbitrary Chern number in chiral twisted graphene multilayers. *Phys. Rev. Lett.* **128**, 176404 (2022).
53. J. P. Eisenstein, K. B. Cooper, L. N. Pfeiffer, K. W. West, Insulating and fractional quantum Hall states in the first excited Landau level. *Phys. Rev. Lett.* **88**, 076801 (2002).
54. S. Chen *et al.*, Competing fractional quantum Hall and electron solid phases in graphene. *Phys. Rev. Lett.* **122**, 026802 (2019).
55. A. S. Patri, Z. Dong, T. Senthil, Extended quantum anomalous Hall effect in moiré structures: Phase transitions and transport. *Phys. Rev. B* **110**, 245115 (2024).
56. B. Tanatar, D. M. Ceperley, Ground state of the two-dimensional electron gas. *Phys. Rev. B* **39**, 5005-5016 (1989).
57. E. Y. Andrei *et al.*, Observation of a magnetically induced Wigner solid. *Phys. Rev. Lett.* **60**, 2765-2768 (1988).
58. Z. Ma *et al.*, Moiré flat bands of twisted few-layer graphite. *Front. Phys.* **18**, 13307

(2022).

59. V. T. Phong, C. Lewandowski, Coulomb Interaction-stabilized isolated narrow bands with Chern numbers in twisted rhombohedral trilayer-bilayer graphene. arXiv:2505.07981 (2025).
60. C. N. Lau, M. W. Bockrath, K. F. Mak, F. Zhang, Reproducibility in the fabrication and physics of moiré materials. *Nature* **602**, 41-50 (2022).
61. S. Slizovskiy, E. McCann, M. Koshino, V. I. Fal'ko, Films of rhombohedral graphite as two-dimensional topological semimetals. *Commun. Phys.* **2**, 164 (2019).
62. C. Wu, D. Bergman, L. Balents, S. Das Sarma, Flat bands and Wigner crystallization in the honeycomb optical lattice. *Phys. Rev. Lett.* **99**, 070401 (2007).
63. C. Wu, Orbital analogue of the quantum anomalous Hall effect in *p*-band systems. *Phys. Rev. Lett.* **101**, 186807 (2008).
64. R. Bistritzer, A. H. MacDonald, Moire bands in twisted double-layer graphene. *Proc. Natl. Acad. Sci. U.S.A.* **108**, 12233-12237 (2011).
65. A. Sterdyniak, C. Repellin, B. A. Bernevig, N. Regnault, Series of Abelian and non-Abelian states in C>1 fractional Chern insulators. *Phys. Rev. B* **87**, 205137 (2013).

**Acknowledgements**
We thank Chao Zhang from the Instrumentation and Service Center for Physical Sciences (ISCPS) at Westlake University for technical and facility support in data acquisition. We also thank the Instrumentation and Service Center for Molecular Sciences (ISCMS) at Westlake University for facility support.
**Funding:** This work was funded by National Natural Science Foundation of China (Grant No. 12274354, S.X.; No. 22473099, N.X.), the Zhejiang Provincial Natural Science Foundation of China (Grant No. LR24A040003, S.X.; No. XHD23A2001, S.X.), and Westlake Education Foundation at Westlake University. K.W. and T.T. acknowledge support from the JSPS KAKENHI (Grant Nos. 21H05233 and 23H02052) and World Premier International Research Center Initiative (WPI), MEXT, Japan.
**Author contributions:** S.X. and N.X. conceived the idea and supervised the project. N.L. fabricated the devices with the assistance of Z.C., W.Z., H.X., X.F., L.W., X.Z., and Q.C.. N.L. performed transport measurements with the assistance of J.D. and L.Z.. N.L., N.X., and S.X. performed data analysis. K.W. and T.T. grew h-BN crystals. N.L., N.X., and S.X. wrote the manuscript. All authors contributed to the discussions.
**Competing interests:** The authors declare no competing financial interests.
**Data and materials availability:** The data shown in the main figures and other findings that support this study are available from the corresponding authors upon reasonable request.

**Supplementary Materials**
Materials and Methods
Supplementary Text
Figs. S1 to S12
Tables S1

Figures

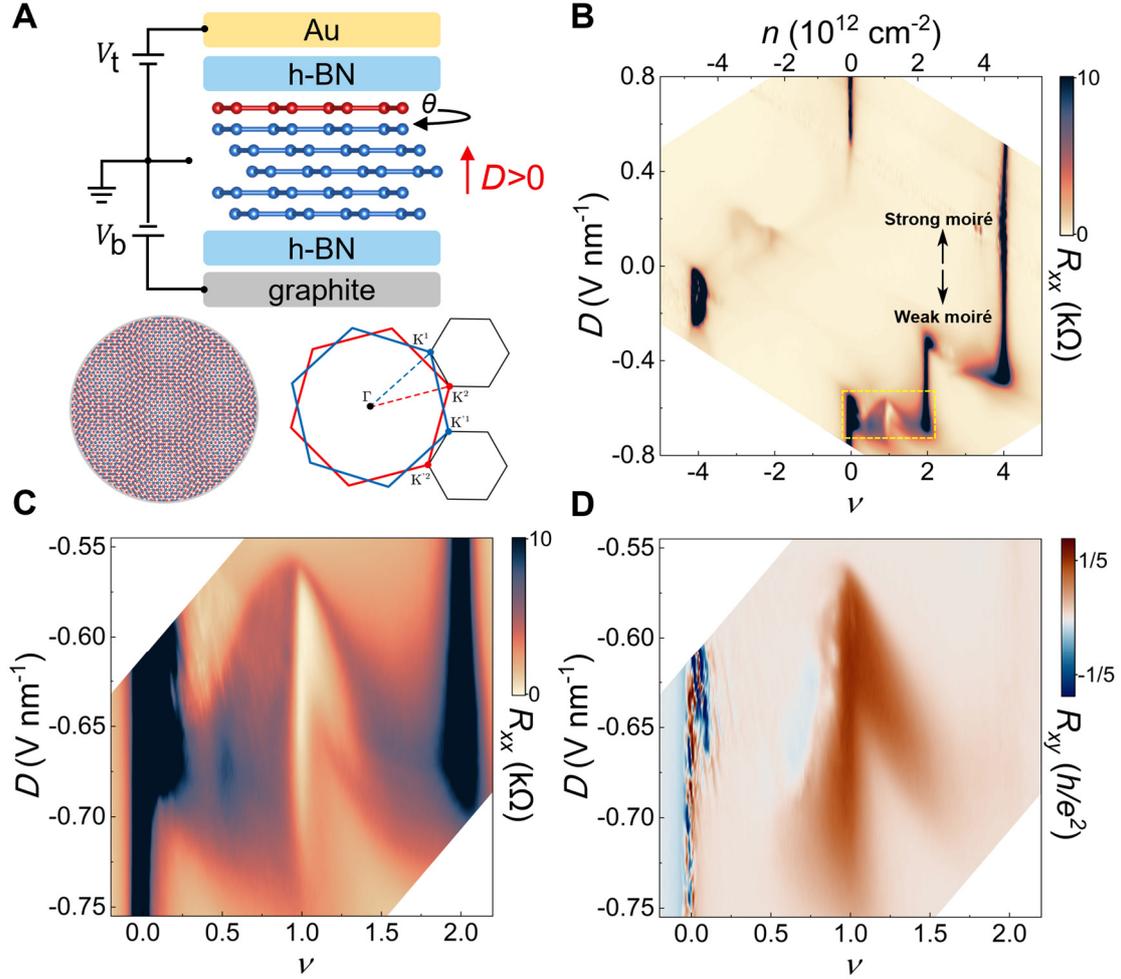

**Fig. 1. Phase diagram of twisted monolayer-rhombohedral pentalayer graphene with a twist angle of $\theta \approx 1.40°$ (device D1).** (**A**) Top panel: schematic of dual-gate structure for twisted graphene devices. Bottom panel: illustration of the moiré superlattice formed at the interface between monolayer graphene and rhombohedral pentalayer graphene. The mini-Brillouin zone is constructed from the twist between the two K (K') wavevectors for the two layers. (**B**) Full map of longitudinal resistance $R_{xx}$ as a function of moiré filling factor $\nu$ (or carrier density $n$ on top $x$-axis) and applied displacement field $D$. Electron carriers accumulate at weak moiré interface for $D < 0$, as illustrated. The data were taken at $T = 50$ mK and $B = 0$ T. (**C** and **D**), Fine maps of symmetrized $R_{xx}$ (C) and anti-symmetrized Hall resistance $R_{xy}$ (D) as a function of $\nu$ and $D$ in the range marked by yellow dashed box in (B). The data were measured at $T = 50$ mK and $B = \pm 0.1$ T.

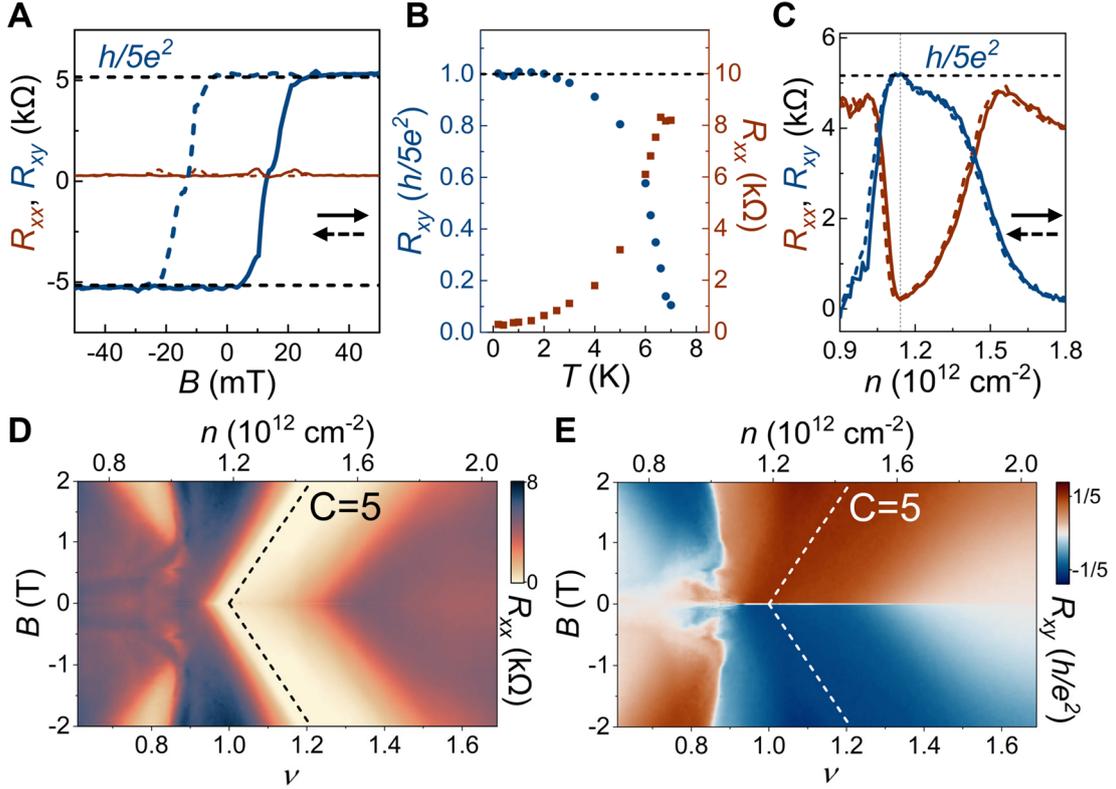

**Fig. 2. Quantum anomalous Hall insulators at $\nu = 1$ (device D1).** (**A**) Magnetic hysteresis loop taken by measuring $R_{xx}$ (red curves) and $R_{xy}$ (blue curves) while sweeping $B$ forward (solid lines) and backward (dashed lines). (**B**) Temperature dependence of $R_{xx}$ and $R_{xy}$ obtained at a small $B = 50$ mT. The data in (A) and (B) were taken at a fixed $\nu = 1$ and $D = -0.620$ V nm$^{-1}$. (**C**) Measurement of symmetrized $R_{xx}$ (red curves) and anti-symmetrized $R_{xy}$ (blue curves) as a function of $n$ at a fixed $D = -0.620$ V nm$^{-1}$ and $B = \pm 50$ mT. The solid (dashed) lines represent scans of $n$ forward (backward). (**D** and **E**) Fan diagrams of $R_{xx}$ (D) and $R_{xy}$ (E) as a function of $B$ and $\nu$ at a fixed $D = -0.620$ V nm$^{-1}$. Dashed lines identify the expected dispersions based on the Streda formula, from which a Chern number of $C = 5$ is extracted.

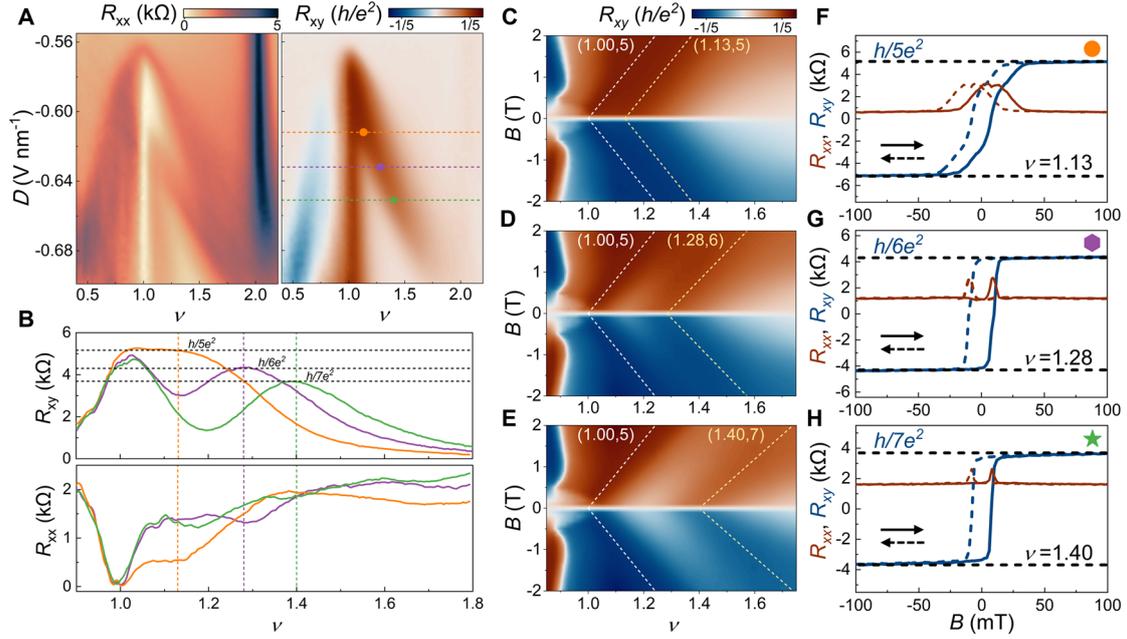

**Fig. 3. Incommensurate quantum anomalous Hall insulators at $1 < \nu < 1.5$ (device D2).** (**A**) Phase diagram near $\nu = 1$ on $D < 0$ side of device D2 with a twist angle of $\theta \approx 1.32°$. The color maps show symmetrized $R_{xx}$ (left panel) and anti-symmetrized $R_{xy}$ (right panel) as a function of $\nu$ and $D$ measured at $B = \pm 0.1$ T. (**B**) $R_{xy}$ and corresponding $R_{xx}$ as a function of $\nu$ extracted along the orange, purple and green dashed lines in (A). (**C** to **E**) Fan diagram of anti-symmetrized $R_{xy}$ as functions of $B$ and $\nu$ obtained at fixed $D = -0.612$ V nm$^{-1}$ (C), $D = -0.632$ V nm$^{-1}$ (D), $D = -0.651$ V nm$^{-1}$ (E), marked by the orange, purple, green dashed lines in (A), respectively. White and pale yellow dashed lines show the expected evolution of $(\nu, C)$ based on the Streda formula. (**F** to **H**) Magnetic hysteresis of $R_{xx}$ (red curves) and $R_{xy}$ (blue curves) by sweeping $B$ forward (solid lines) and backward (dashed lines) at fixed $D$ and $\nu$: $D = -0.612$ V nm$^{-1}$ and $\nu = 1.13$ (F), $D = -0.632$ V nm$^{-1}$ and $\nu = 1.28$ (G), $D = -0.651$ V nm$^{-1}$ and $\nu = 1.40$ (H).

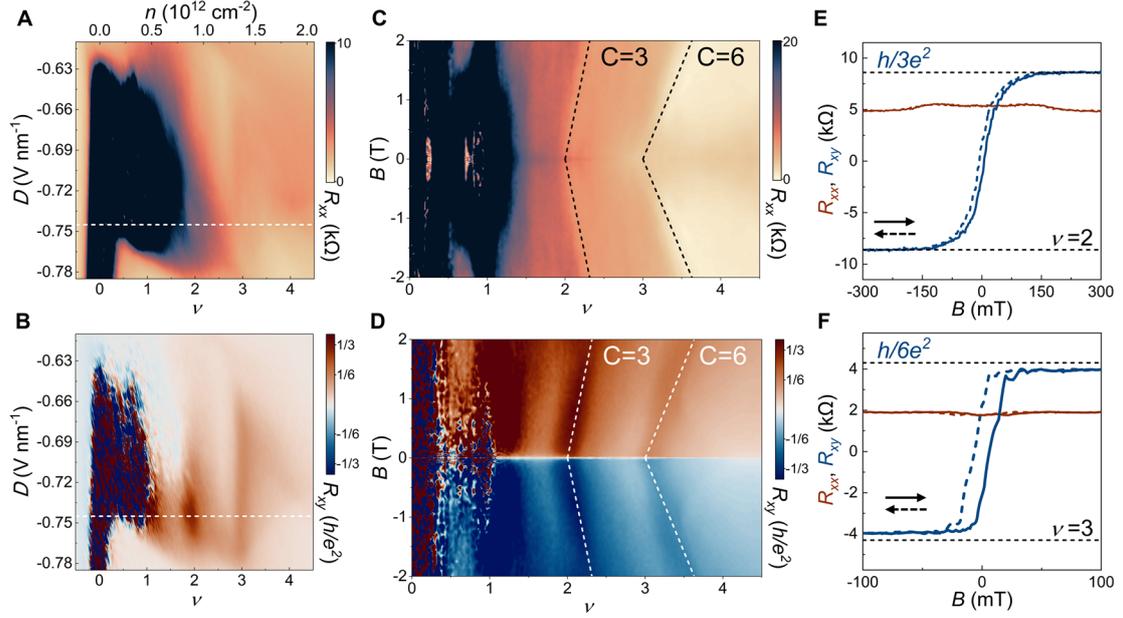

**Fig. 4. High-Chern-number Chern insulators in a device with a twist angle of $\theta \approx 0.89°$ (device D3).** (**A** and **B**) Phase diagrams showing symmetrized $R_{xx}$ (A) and anti-symmetrized $R_{xy}$ (B) as a function of $\nu$ (or $n$) and $D$ measured at $B = \pm 0.1$ T. (**C** and **D**) Fan diagrams of $R_{xx}$ (C) and $R_{xy}$ (D) as a function of $B$ and $\nu$ at a fixed $D = -0.743$ V nm$^{-1}$. Dashed lines identify the expected dispersions based on the Streda formula, from which Chern numbers of $C = 3$ and $C = 6$ are extracted at $\nu = 2$ and $\nu = 3$, respectively. (**E**) Magnetic hysteresis of $R_{xx}$ (red curves) and $R_{xy}$ (blue curves) while sweeping $B$ forward (solid lines) and backward (dashed lines) at fixed $\nu = 2$ and $D = -0.743$ V nm$^{-1}$. (**F**) Magnetic hysteresis of $R_{xx}$ (red curves) and $R_{xy}$ (blue curves) at fixed $\nu = 3$ and $D = -0.732$ V nm$^{-1}$.

## Supplementary Materials

**Materials and Methods**

Device fabrication

Our heterostructures consist of monolayer graphene and rhombohedral pentalayer graphene encapsulated between top and bottom h-BN, with a graphite back gate. They were assembled via the "cut-and-stack" technique using a standard dry-transfer process. Graphene flakes with monolayer-pentalayer steps, as well as h-BN flakes, were mechanically exfoliated onto a $SiO_2$/Si substrate (with ~300 nm oxide thickness). Rhombohedral domains within the pentalayer graphene were identified and spatially resolved via Raman spectroscopy or infrared imaging (Fig. S1) (*1, 2*). The primary challenges in sample assembly include the instability of rhombohedral domains, relaxation of the twist angle, and strain introduced during the transfer process. To mitigate these issues and improve fabrication yield, graphene flakes were deliberately cut into two isolated pieces using an atomic force microscope (AFM). The flakes were shaped into rectangular shapes with an array of finger-like edge patterns to suppress angular relaxation during stacking. A poly(bisphenol A carbonate) (PC)/polydimethylsiloxane (PDMS) stamp was employed to sequentially pick up the top h-BN, monolayer graphene, and rhombohedral pentalayer graphene. The twist angle between two graphene pieces was targeted by precisely rotating the sample stage during transfer. The whole stack was then released onto a preassembled h-BN/bottom graphite substrate that had been cleaned by AFM. During the transfer, we intentionally avoid the crystallographic alignment between graphene and h-BN. The alive rhombohedral stacking in the final heterostructure was confirmed through a second Raman mapping, as demonstrated in Fig. S2. AFM imaging was further used to identify bubble-free regions for device fabrication.

The heterostructures were patterned into Hall bar geometries using standard electron-beam lithography followed by reactive-ion etching. Electrical contacts and the top gate were deposited by electron-beam evaporation of Cr/Au.

Electrical transport measurements

All transport measurements were performed in a dilution refrigerator (Oxford Triton) with a base temperature of $T = 50$ mK, unless otherwise specified. To minimize electronic noise, low-temperature RC and RF filters (QDevil) were installed at the mixing chamber plate. Standard low-frequency lock-in techniques (Zurich Instruments MFLI) were employed to measure the longitudinal and Hall resistance ($R_{xx}$ and $R_{xy}$) at an excitation frequency of 17.7 Hz. A low excitation current of 1 nA - 5 nA was used to probe fragile topological states. Gate voltages were applied using a Keithley 2614B source-measure unit.

Converting dual-gate map to $n - D$ map

The carrier density $n$ and displacement field $D$ in the devices were calculated from the applied top and bottom gate voltages ($V_t$ and $V_b$) using the relations: $n = \frac{(C_b V_b + C_t V_t)}{e} -$ $n_{\text{offset}}$ and $D = \frac{(C_b V_b - C_t V_t)}{2\varepsilon_0} - D_{\text{offset}}$. Here, $C_b$ and $C_t$ are the capacitances of the bottom and top gates, respectively, calibrated via Landau-level sequences at high $B$ (see Fig. S5). $e$ is the elementary charge and $\varepsilon_0$ is the vacuum permittivity. Offset values $n_{\text{offset}}$ and $D_{\text{offset}}$, which arise from environment-induced carrier doping and electrostatic potential, were inferred from aligning the resistance features in the dual-gate resistance map. The $n$ is converted to the moiré filling factor $\nu$ by $\nu = 4n/n_s$, where $n_s$ is the carrier density corresponding to full filling of a moiré band. Fig. S3 demonstrates the $V_t$ - $V_b$ map and the corresponding converted $\nu - D$ maps.

Symmetrized $R_{xx}$ and antisymmetrized $R_{xy}$

The measured Hall resistance $R_{xy}$ inevitably includes contributions from the longitudinal resistance $R_{xx}$ due to imperfect Hall bar geometry. To eliminate these components, we employed standard symmetrization and antisymmetrization procedures. For data measured at a fixed $B$, we calculated the antisymmetrized Hall resistance and symmetrized longitudinal resistance using: $R_{xy}(\pm B) = [R_{xy}^{\text{raw}}(+B) - R_{xy}^{\text{raw}}(-B)]/2$ and $R_{xx}(\pm B) = [R_{xx}^{\text{raw}}(+B) + R_{xx}^{\text{raw}}(-B)]/2$. For magnetic hysteresis measurements, we employed $R_{xy}^{\text{anti-sym}}(B, \leftarrow) = [R_{xy}^{\text{raw}}(B, \leftarrow) - R_{xy}^{\text{raw}}(-B, \rightarrow)]/2$, $R_{xy}^{\text{anti-sym}}(B, \rightarrow) = [R_{xy}^{\text{raw}}(B, \rightarrow) - R_{xy}^{\text{raw}}(-B, \leftarrow)]/2$, $R_{xx}^{\text{sym}}(B, \leftarrow) = [R_{xx}^{\text{raw}}(B, \leftarrow) + R_{xx}^{\text{raw}}(-B, \rightarrow)]/2$, and $R_{xx}^{\text{sym}}(B, \rightarrow) = [R_{xx}^{\text{raw}}(B, \rightarrow) + R_{xx}^{\text{raw}}(-B, \leftarrow)]/2$. Here, the arrows ($\leftarrow, \rightarrow$) represent the direction of the magnetic-field sweep during measurement.

Twist angle estimation

The phase diagram, such as that shown in Fig. 1B, was used to estimate the twist angle in the measured device. The carrier density corresponding to full filling of a moiré band ($\nu = 4$) is given by $n_s \approx 8\theta^2/\sqrt{3}a^2$, where $a = 0.246$ nm is the lattice constant of graphene. In device D1, we find $n_s = 4.58 \times 10^{12}$ cm$^{-2}$, yielding a twist angle of $\theta \approx 1.40°$.

Alternatively, the twist angle can also be extracted directly from Brown-Zak oscillations. The minima in $R_{xx}$ observed in Brown-Zak oscillations occur at $B = \phi_0/qA$, where $\phi_0 = h/e$ is the magnetic flux quantum, $q$ is an integer, and $A = \sqrt{3}\lambda^2/2$ is the unit cell area of the moiré superlattice. By linearly fitting a plot of $q - 1/B$, we can extract the moiré wavelength $\lambda = 10.14$ nm in device D1 (Fig. S4). This corresponds to a twist angle of $\theta \approx 1.39°$, consistent with the value estimated from the phase diagram.

The estimated twist angles of other devices are summarized in Table S1.

**Supplementary Text**

<u>Trivially topological states at $D > 0$</u>

Similar to the observations in rhombohedral graphene/h-BN moiré superlattices (*3-5*), we find that the topologically nontrivial states in our system only emerge when electrons are polarized toward moiréless interfaces, as confirmed by Fig. S3. Evern more surprisingly, no correlated states are observed when electrons are polarized toward moiré interfaces ($D > 0$) as shown in Fig. S3D, while the full filling ($\nu = 4$) gaps are strong. Correlated states are expected to occur at partial fillings of moiré bands (such as $\nu = 1, 2, 3$), and indeed such states have been observed in rhombohedral graphene/h-BN moiré superlattices (*3-5*). Under applied magnetic fields, correlated states at $\nu = 2$ can emerge; however, they are topologically trivial, as indicated by their non-dispersive Landau fan diagrams and the absence of anomalous Hall resistance (see Fig. S6). These unconventional behaviors call for future theoretical studies.

<u>Switching Chern states at finite magnetic fields</u>

The topological states in our system can be tuned not only by displacement fields, as demonstrated in Fig. 3, but also by magnetic fields. Fig. S7 displays a series of topological phase transitions occurring at moderate magnetic fields. Specifically, we observe that the $C = 5$ state switches to a $C = 4$ state at $B \approx 1.6$ T, followed by a further switching to a $C = 3$ state at $B \approx 4.8$ T. Notably, the emergence of the $C = 4$ and $C = 3$ states is unlikely to originate from conventional Landau level quantization, as they solely appear. Indeed, at higher magnetic fields, Landau level sequences emanating from $\nu = 1$ and $\nu = 0$ emerge collectively, consistent with standard quantum Hall behavior.

Additionally, we find that the charity of the $C = 5$ state at $\nu = 1$ can be switched under applied small magnetic fields. As shown in Fig. S8, the $C = -5$ state coexists with the $C = 5$ state at $B > 0.5$ T. Sweeping $\nu$ at a fixed $B = 2$ T, we observe that the sign of the anomalous $R_{xy}$ is reversible for $\nu > 1$ and $\nu < 1$. However, unlike twisted Bernal graphene systems (*6*), no hysteresis is detected upon sweeping $\nu$. This phenomenon resembles the Chern states reported in rhombohedral multilayer graphene/h-BN moiré superlattices (*4, 5*).

Notably, this switchable chirality feature has been observed in device D1, D2, and D4 with similar twist angles, but is absent in device D3 with a smaller twist angle, indicating that the topological phase transitions in our system are highly sensitive to the twist angle.

In summary, we demonstrate that the high-Chern-number Chern states in our system are highly tunable. Both their absolute value and sign can be switched by either electric fields or magnetic fields.

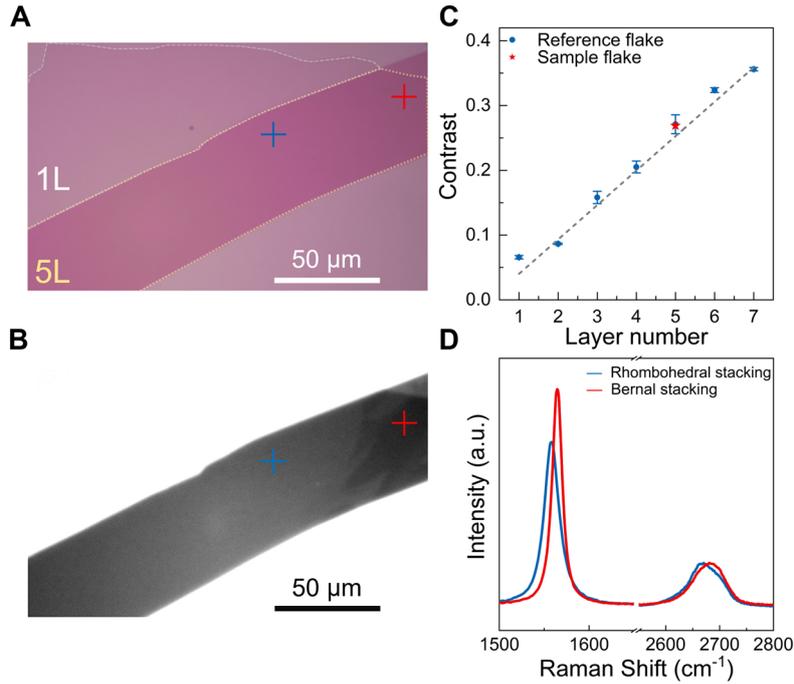

**Fig. S1. Identification of layer number and rhombohedral domains in pristine graphene flake.** (**A**) Optical image of a pentalayer graphene with a connected monolayer graphene. (**B**) Infrared image of the same flake as in (A) taken by an InGaAs camera. Two different contrasts can be observed with the bright (black) one corresponding to rhombohedral (Bernal) stacking. (**C**) Optical contrast as a function of layer number. The red star marks the pentalayer graphene used for the device fabrication. (**D**) Raman spectroscopy taken from the points marked by blue and red crosses in (A) and (B). The blue and red curves confirm the rhombohedral- and Bernal-stacking domains, respectively.

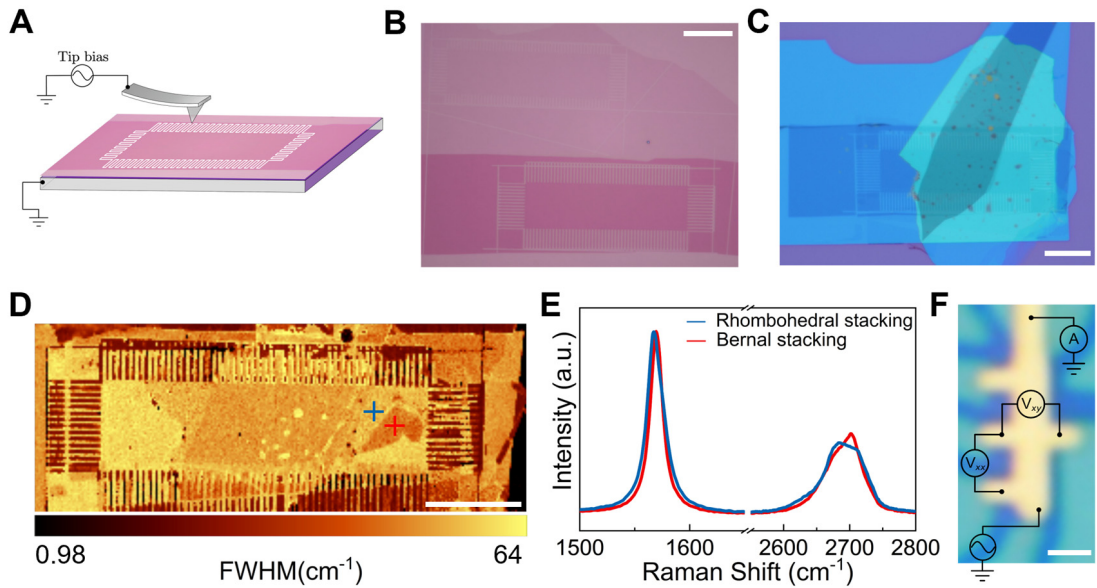

**Fig. S2. Fabrication process of monolayer-rhombohedral pentalayer graphene devices.** (**A**) Schematic of cutting process using an AFM tip. The inset shows finger-like pattern near the edge. (**B**) Optical image of the cut flake. Monolayer graphene and rhombohedral pentalayer graphene with finger-like edges are isolated from each other and Bernal-stacking domains. (**C**) Optical image of the final stack consisting of bottom graphite, bottom h-BN, rhombohedral pentalayer graphene, monolayer graphene, and top h-BN. (**D**) Raman map of the final stack taken by plotting full width at half maximum of 2D Raman peaks. (**E**) Individual Raman spectra taken from the positions marked by blue and red crosses in (D). The blue and red curves represent rhombohedral- and Bernal-stacking domains in the final stacks, respectively. (**F**) Optical image of a final device (device D1). The measuring configurations of longitudinal and Hall resistances are marked. Scale bars in (B-D) are 15 μm. Scale bar in (F) is 1 μm.

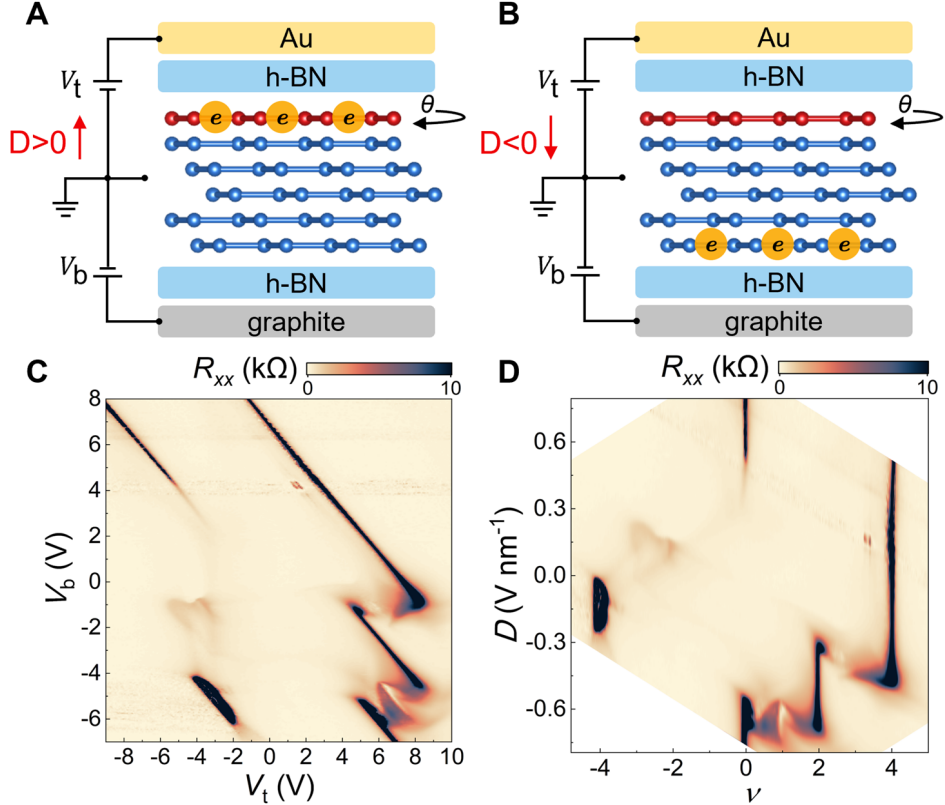

**Fig. S3. Layer polarization of electrons under defined displacement fields.** (**A-B**) Schematics of the layer polarization of electrons in the low-energy conduction band under $D > 0$ (A) and $D < 0$ (B). (**C**) Raw data plotted by measured $R_{xx}$ as functions of top gate ($V_t$) and bottom gate ($V_b$) at $T = 50$ mK and $B = 0$ T. (**D**) Converted $\nu-D$ map of $R_{xx}(\nu, D)$ from the data of $R_{xx}(V_t, V_b)$ in (C). The results confirm that topological nontrivial states discussed in the main text emerge under $D < 0$, when electrons are polarized far away from the moiré interface.

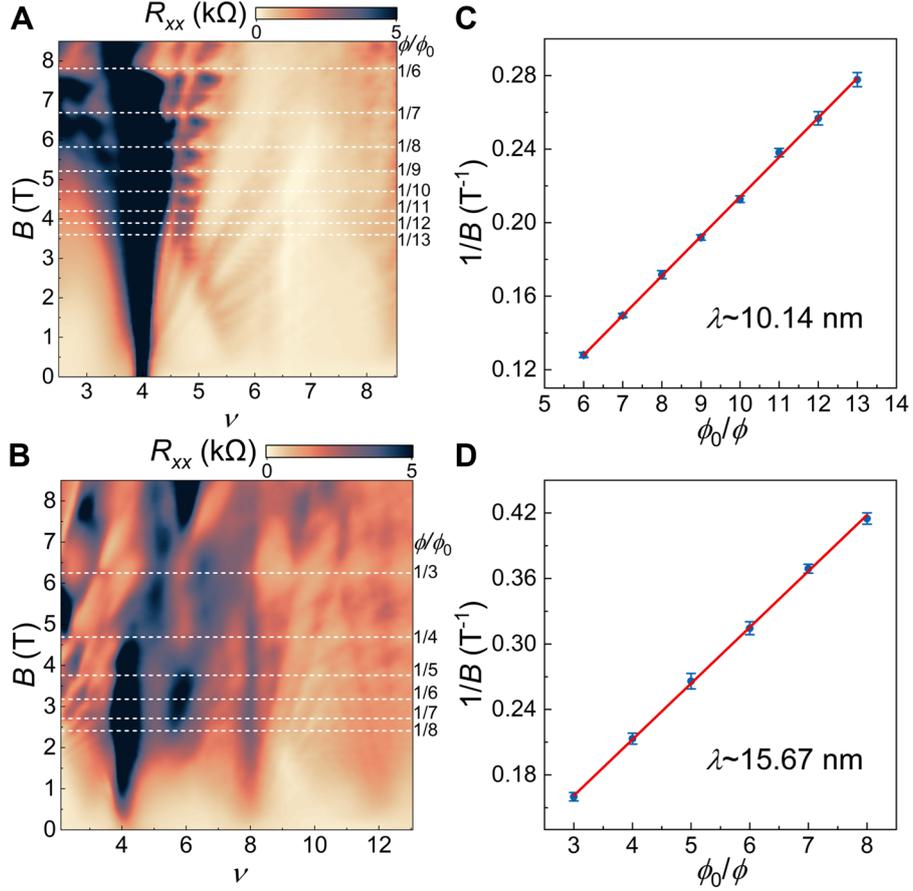

**Fig. S4. Brown-Zak oscillations and moiré wavelength determination.** (**A-B**) Landau fan diagrams of $R_{xx}$ as functions of $v$ and $B$ for two representative devices measured at $T = 50$ mK: device D1 (A), device D3 (B). The dashed horizontal lines mark minimum $R_{xx}$ in Brown-Zak oscillations when the magnetic flux through the moiré unit cell is commensurate with the flux quantum, i.e., $\phi = \phi_0/q$ with an integer $q$. Data in (A) and (B) are taken at fixed $D = -0.190$ V nm$^{-1}$ and $D = 0.390$ V nm$^{-1}$, respectively. (**C-D**) Linear fit of the oscillation period. By fitting $1/B$ values at the minimum $R_{xx}$ in Brow-Zak oscillations as a function of $q = \phi_0/\phi$, the moiré wavelengths are determined to be 10.14 nm (C) and 15.67 nm (D) corresponding to the twist angles of $\theta \approx 1.39°$ and $\theta \approx 0.90°$ for device D1 and device D3, respectively.

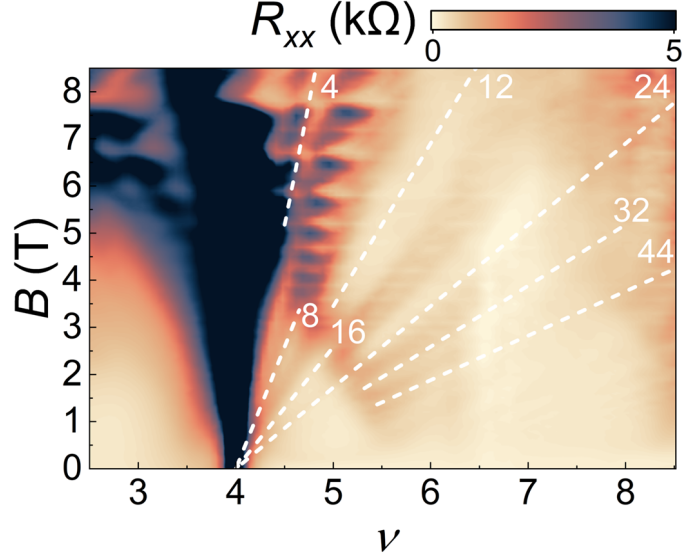

**Fig. S5. Landau level sequences used for carrier density calibration.** Landau fan diagram of $R_{xx}$ as functions of $v$ and $B$ measured at a fixed $D = -0.190$ V nm$^{-1}$ and $T = 50$ mK (device D1). The white dashed lines show the trajectories of quantum Hall states with Landau level fillings of $v' = 4, 8, 12, ....$ The positions of these states are used to calibrate the carrier density.

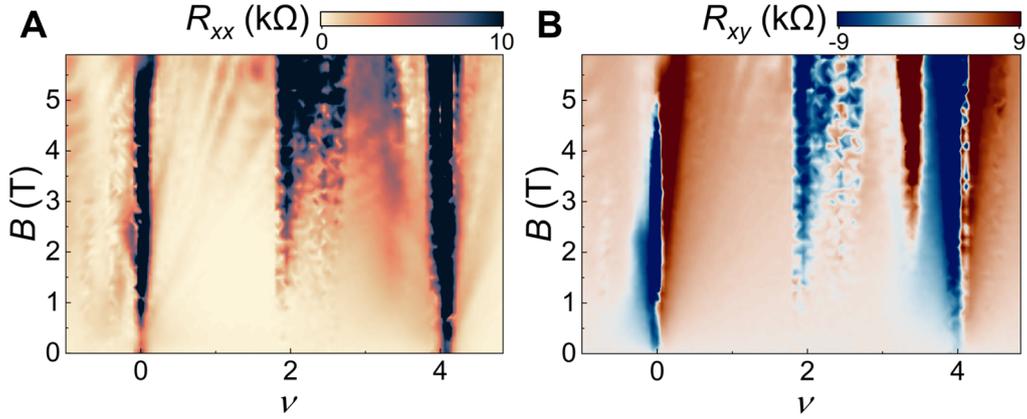

**Fig. S6. Topologically trivial states on $D > 0$ side (device D1).** (A-B) Landau fan diagrams of $R_{xx}$ (A) and $R_{xy}$ (B) as functions of $v$ and $B$ measured at a fixed $D = 0.530$ V nm$^{-1}$ and $T = 50$ mK. A correlated resistive state emerges at $v = 2$ when $B > 1$ T, whose position is independent of $B$. Both the non-dispersive Landau fan diagrams and the absence of anomalous Hall resistance near $B = 0$ indicate the correlated state at $v = 2$ is a topologically trivial state.

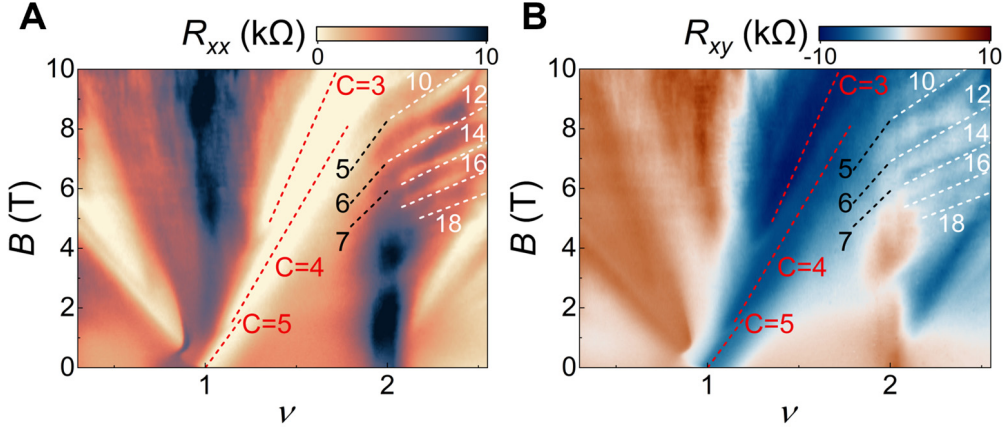

**Fig. S7. Topological phase transition at high magnetic fields (device D2).** (A-B) Landau fan diagrams of $R_{xx}$ (A) and $R_{xy}$ (B) as functions of $\nu$ and $B$ measured at a fixed $D = -0.586$ V nm$^{-1}$ and $T = 50$ mK. The white (red) dashed lines represent Landau level sequences emanating from $\nu = 0$ ($\nu = 1$). The orange dashed lines mark the Chern insulators emanating from $\nu = 1$. Magnetic fields induced topological phase transitions can be observed from $C = 5$ to $C = 4$ Chern states at $B \approx 1.6$ T and from $C = 4$ to $C = 3$ Chern states at $B \approx 4.8$ T.

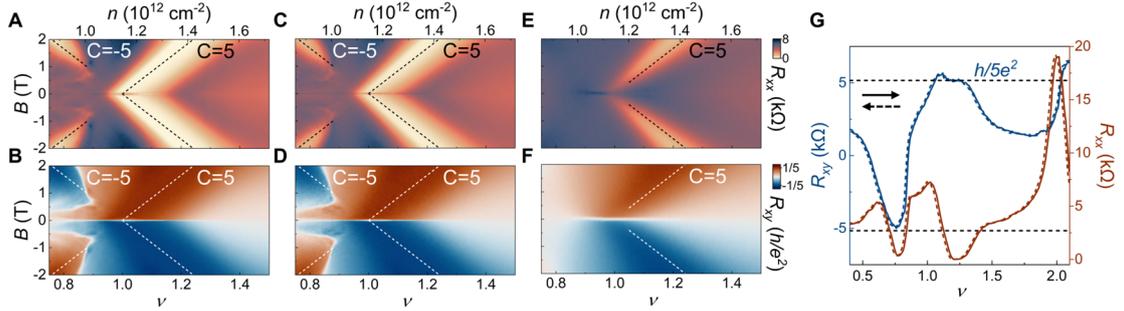

**Fig. S8. Chirality switching of $C = 5$ states at small magnetic fields and their temperature dependence (device D1).** (A-B) Fan diagrams plotted by $R_{xx}$ (A) and $R_{xy}$ (B) as functions of $\nu$ and $B$ measured at a fixed $D = -0.600$ V nm$^{-1}$ and $T = 50$ mK. (C-F) Similar measurements at different temperatures of $T = 2$ K (C&D) and $T = 5$ K (E&F). (G) Measurement of symmetrized $R_{xx}$ (red curves) and anti-symmetrized $R_{xy}$ (blue curves) as a function of $n$ at a fixed $D = -0.600$ V nm$^{-1}$ and $B = \pm 2$ T. The solid (dashed) lines represent scans of $n$ forward (backward).

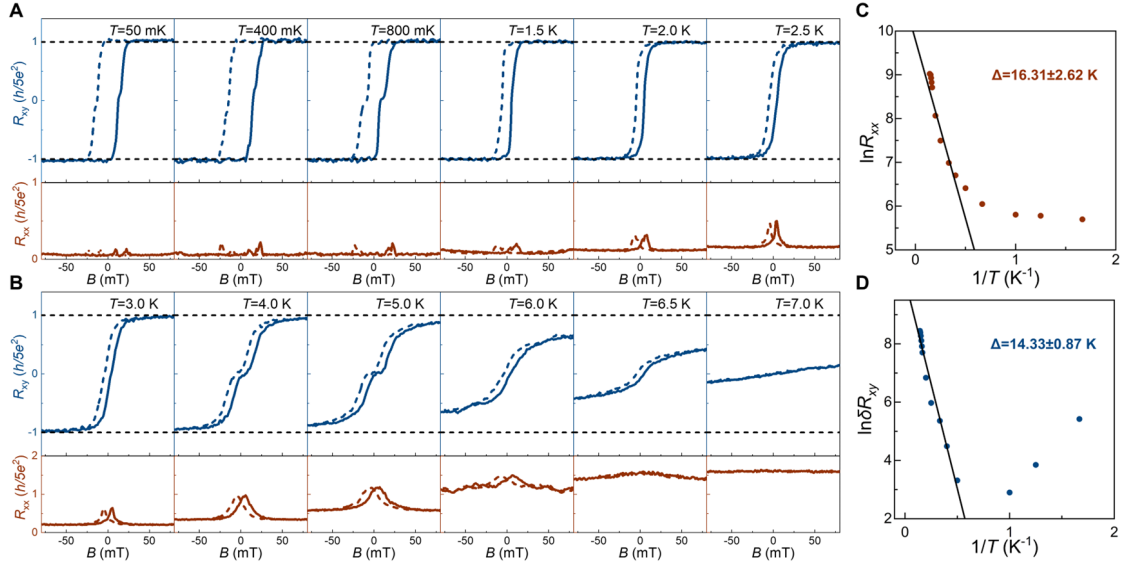

**Fig. S9. Temperature dependence of quantum anomalous Hall states with $C = 5$ (device D1).** (**A-B**) Magnetic hysteresis loops measured by sweeping $B$ forward and backward at various temperatures: from $T = 50$ mK to $T = 2.5$ K (A); from $T = 3.0$ K to $T = 7.0$ K (B). The upper and lower panels display $R_{xy}$ and $R_{xx}$ as a function of $B$, respectively. The data were measured at $\nu = 1$ and $D = -0.620$ V nm$^{-1}$. (**C-D**) Arrhenius plots of $R_{xx}$ (C) and the deviation of $R_{xy}$ from its quantized value denoted as $\delta R_{xy} = \frac{h}{5e^2} - |R_{xy}|$ (D) as a function of $T^{-1}$. The data are obtained at $B = \pm 50$ mT. The thermal activation gaps were derived using the exponential relations $R_{xx} \propto e^{-\Delta/2k_BT}$ and $\delta R_{xy} \propto e^{-\Delta/k_BT}$, yielding gap sizes of $16.31 \pm 2.62$ K and $14.33 \pm 0.87$ K, respectively.

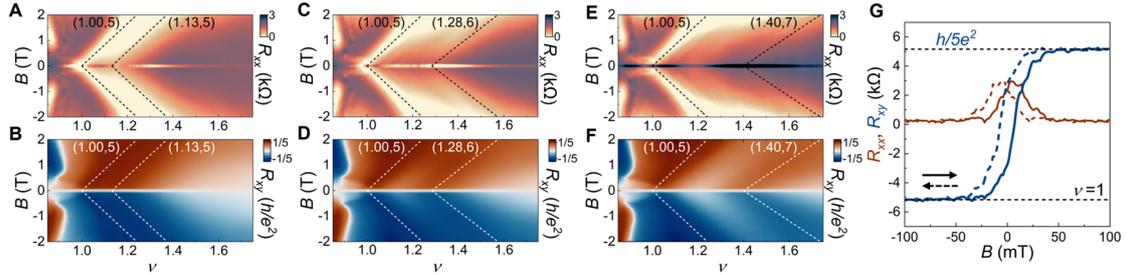

**Fig. S10. Additional data for device D2.** (**A-F**) Landau fan diagram taken at different $D$. (A, C, E) and (B, D, F) show $R_{xx}$ and $R_{xy}$ as functions of $\nu$ and $B$, respectively. The $D$ used are $-0.612$ V nm$^{-1}$ (A&B); $-0.632$ V nm$^{-1}$ (C&D); $-0.651$ V nm$^{-1}$ (E&F). (**G**) Magnetic hysteresis taken at $\nu = 1$ and $D = -0.605$ V nm$^{-1}$.

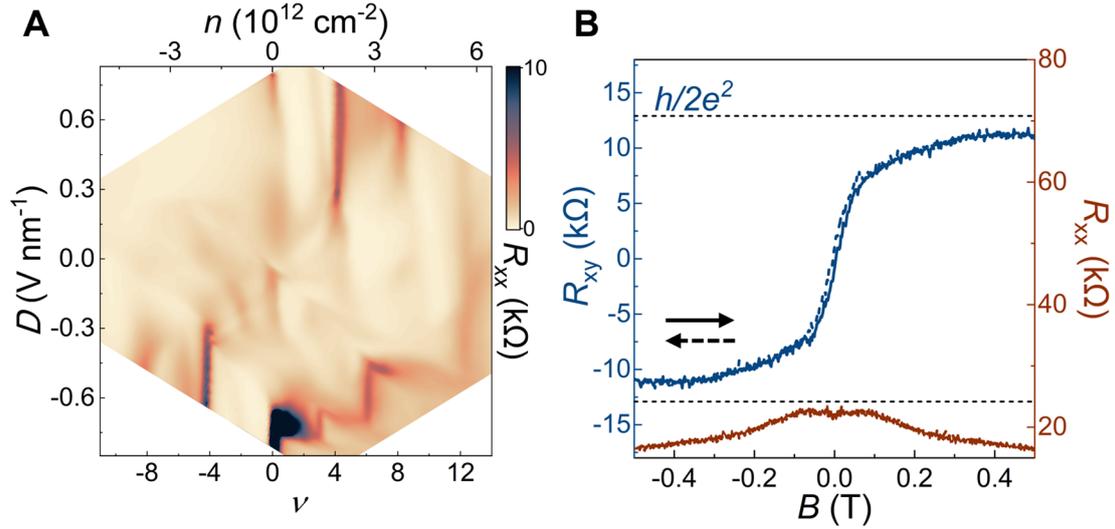

**Fig. S11. Full map of device D3 and magnetic hysteresis near $\nu = 1$.** (A) $\nu - D$ map of $R_{xx}$ measured at $T = 50$ mK. A small $B = 1$ T was applied in the measurement in order to enhance the visualization of the gapped states. (B) Magnetic hysteretic loop at fixed $\nu = 1.2$ and $D = -0.743$ V nm$^{-1}$ by sweeping $B$ forward and backward.

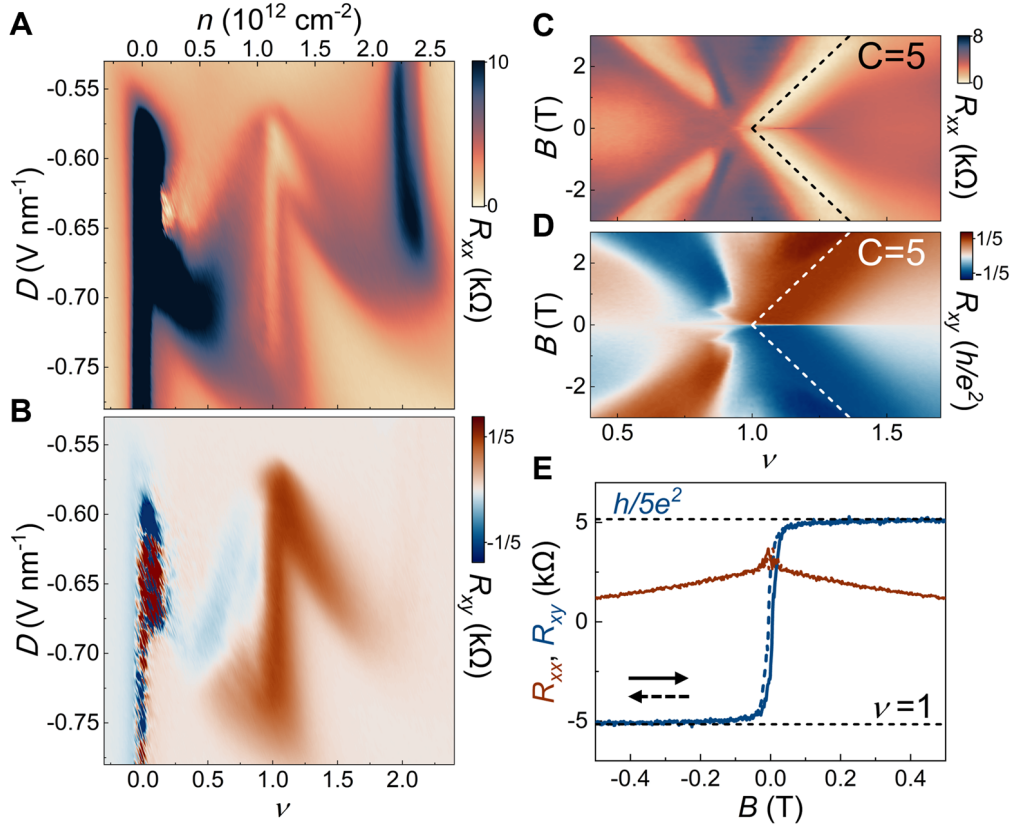

**Fig. S12. Chern insulators in device D4 with a twist angle of $\theta \approx 1.32°$.** (A-B) Fine maps of symmetrized $R_{xx}$ (A) and anti-symmetrized Hall resistance $R_{xy}$ (B) as a function of $\nu$ and $D$ near $\nu = 1$. The data were measured at $T = 50$ mK and $B = \pm 0.1$ T. (C-D) Fan diagrams of $R_{xx}$ (C) and $R_{xy}$ (D) as a function of $B$ and $\nu$ at a fixed $D = -0.600$ V nm$^{-1}$. Dashed lines identify the expected dispersions of $C = 5$ Chern states based on the Streda formula. (E) Magnetic hysteretic loop of $R_{xx}$ (red curves) and $R_{xy}$ (blue curves) taken at $\nu = 1$ and $D = -0.600$ V nm$^{-1}$.

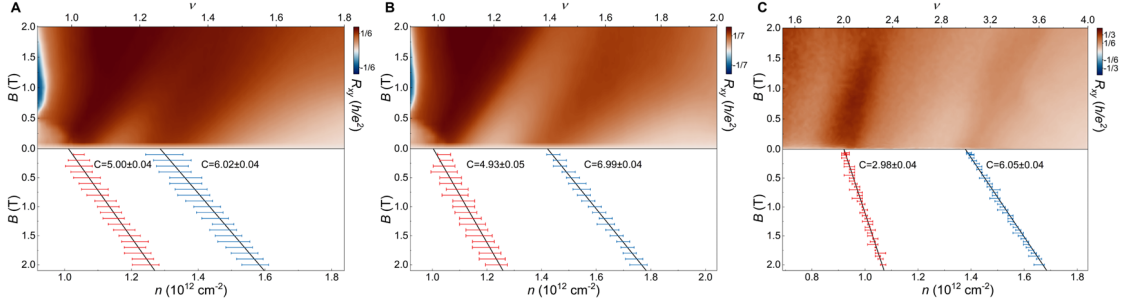

**Fig. S13. Chern number determination.** To quantitatively extract the Chern number from fan diagrams according to the Streda formula, we plot the local maxima in $R_{xy}$ as a function of $\nu$ for each magnetic field. The error bars represent a 5% tolerance threshold in $R_{xy}$ of their local maxima. Black lines illustrate the linear fittings using the Streda formula $\frac{\partial n}{\partial B} = C \frac{e}{h}$, with the fitted Chern numbers indicated in the figures. (**A-B**) Fan diagrams and corresponding fitting results for device D2 measured at fixed $D = -0.632$ V nm$^{-1}$ (A) and $D = -0.651$ V nm$^{-1}$ (B). (**C**) Similar data for device D3 measured at fixed $D = -0.743$ V nm$^{-1}$. The fitted Chern numbers are consistent with the measured quantized Hall resistances in corresponding magnetic hysteresis loops shown in Fig. 3 and Fig. 4.

**Table S1. Summary of measured devices.**

| Device | Twist angle determined by full filling(°) | Moiré wavelength determined by full filling(nm) | Twist angle determined by Brown-Zak oscillations(°) | Moiré wavelength determined by Brown-Zak oscillations(nm) |
|---|---|---|---|---|
| D1 | 1.40 | 10.07 | 1.39 | 10.14 |
| D2 | 1.32 | 10.68 | 1.34 | 10.52 |
| D3 | 0.89 | 15.84 | 0.90 | 15.67 |
| D4 | 1.32 | 10.68 | 1.32 | 10.68 |

**References**
1. Z. Feng *et al.*, Rapid infrared imaging of rhombohedral graphene. *Phys. Rev. Applied* **23**, 034012 (2025).
2. Z. Lu *et al.*, Extended quantum anomalous Hall states in graphene/hBN moiré superlattices. *Nature* **637**, 1090–1095 (2025).
3. Z. Lu *et al.*, Fractional quantum anomalous Hall effect in multilayer graphene. *Nature* **626**, 759-764 (2024).
4. J. Ding *et al.*, Electric-field switchable chirality in rhombohedral graphene Chern insulators stabilized by tungsten diselenide. *Phys. Rev. X* **15**, 011052 (2025).
5. J. Xie *et al.*, Tunable fractional Chern insulators in rhombohedral graphene superlattices. *Nat. Mater.* **24**, 1042-1048 (2025).
6. H. Polshyn *et al.*, Electrical switching of magnetic order in an orbital Chern insulator. *Nature* **588**, 66-70 (2020).